\documentclass[11pt,twoside]{article}
\usepackage{asp2010}
\usepackage{graphicx}

\def\ts     {\thinspace}
\def\kms    {\ifmmode{{\rm \ts km\ts s}^{-1}}\else{\ts km\ts s$^{-1}$}\fi}
\def\msol   {\ifmmode{{\rm M}_{\odot}}\else{M$_{\odot}$}\fi}
\def\lsol   {\ifmmode{{\rm L}_{\odot}}\else{L$_{\odot}$}\fi}
\def\zsol   {\ifmmode{{\rm Z}_{\odot}}\else{Z$_{\odot}$}\fi}
\def\etal   {{\rm et\ts al. }}

\resetcounters

\begin{document}

\title{Polar rings and the 3D-shape of dark matter}
\author{Francoise Combes}
\affil{Observatoire de Paris, LERMA, CNRS, 61 Av de l'Observatoire, 75014 Paris, France}

\begin{abstract}
Polar ring galaxies (PRG) are unique to give insight on the 3D-shape of dark matter haloes.
Some caveats have prevented to draw clear conclusions in previous works.
Also the formation mechanisms need to be well known.
All available information on the topic is reviewed, and criteria are defined
for an ideal PRG system, in the hope of removing the ambiguities and make
progress in the domain. 
\end{abstract}

\section{Introduction}

Polar ring galaxies are clearly multi-spin systems, with two almost perpendicular spins.
The primary object is usually an early-type galaxy, poor in gas, lenticular or elliptical.
The polar ring is generally younger, akin to late-type galaxies, with
large amount of HI gas, young stars with blue colors. The 
frequency of PRG, once corrected from selection biases and projection effects,
has been estimated to  $\sim$5\%  (Whitmore \etal 1990).

Recently, the Galaxy Zoo project in the Sloan data-base has been searched
to find new PRG; Finkelman \etal (2012) identified and observed
16 candidates at slightly higher redshift than Whitmore \etal (1990).
PRG at even higher redshifts were identified with HST images
by Reshetnikov (1997).  Moiseev \etal (2011) have built a catalogue
of 275 PRG candidates from the Galaxy Zoo project.

These new candidates are a promise for the determination 
of the 3D-shape of dark matter halos. However, 
to be able to relate the dark matter of the PRG to dark matter
in spiral progenitors, it is necessary to study the PRG formation scenarios.

\subsection{The ideal polar ring galaxy}

The polar  material should have a large extent,
with a large HI gas radius, to measure the rotation curve
as far as possible from the center and probe the dark halo.

The polar ring should be as wide as possible, and 
even become a polar disk, with young stars and 
ionised gas yielding the H$\alpha$ rotation curve.

The mass of the polar system should be small enough
that the PR material can be considered as test particles in the host potential,
and not perturb the shape of the original halo.

\subsection{Diversity of results}

All previous studies (e.g. Whitmore \etal 1987, Sackett \& Sparke 1990, 
Sackett \etal 1994, Reshetnikov \& Combes 1994, 
Combes \& Arnaboldi 1996, Iodice \etal 2003) 
agree only on one conclusion, that PRG are indeed embedded in a dark halo.

But they conclude to very different 3D-shapes:
almost spherical halo for Whitmore \etal (1987), 
flattened along the equatorial plane of the host (Sackett \etal 1994), 
or flattened along the polar-ring plane (Combes \& Arnaboldi 1996). 
The latter was supported for a large number of PRG by Iodice \etal (2003),
through a study of the Tully-Fisher diagram.

\section{Formation of polar rings}

At least three formation mechanisms have been proposed: 
(1) the accretion scenario, where two interacting galaxies exchange mass (Schweizer \etal 1983; 
Reshetnikov \& Sotnikova 1997); 
(2) the merging scenario (Bekki 1997, 1998; Bournaud \& Combes 2003); 
(3) the cosmic formation scenario, where the PRG form through the misaligned accretion 
of gas from cosmic filaments (Maccio \etal 2006; Brook \etal 2008). 

\begin{figure*}[ht]
\centerline{
\includegraphics[angle=-0,width=6cm]{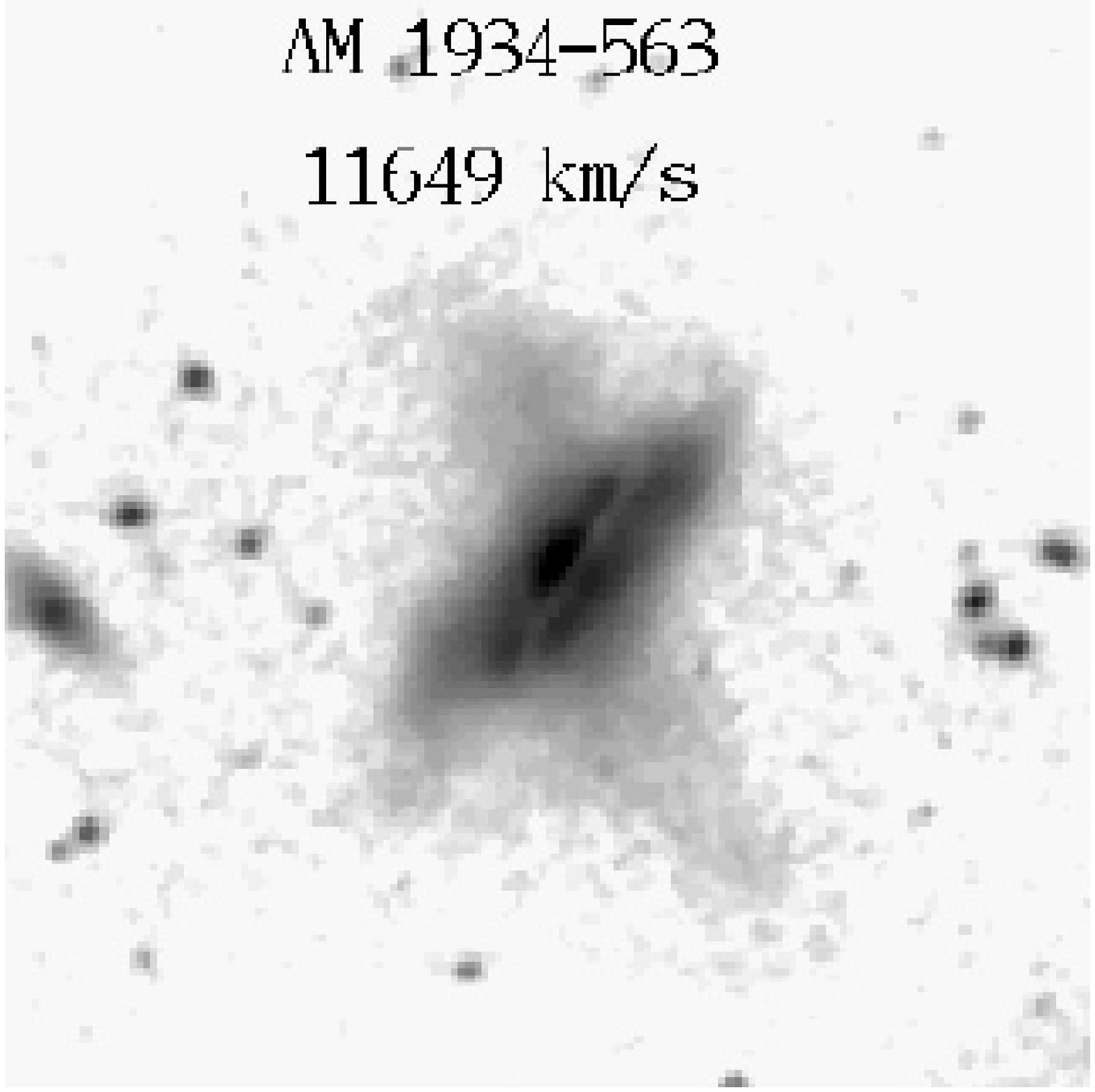}
\includegraphics[angle=-0,width=6cm]{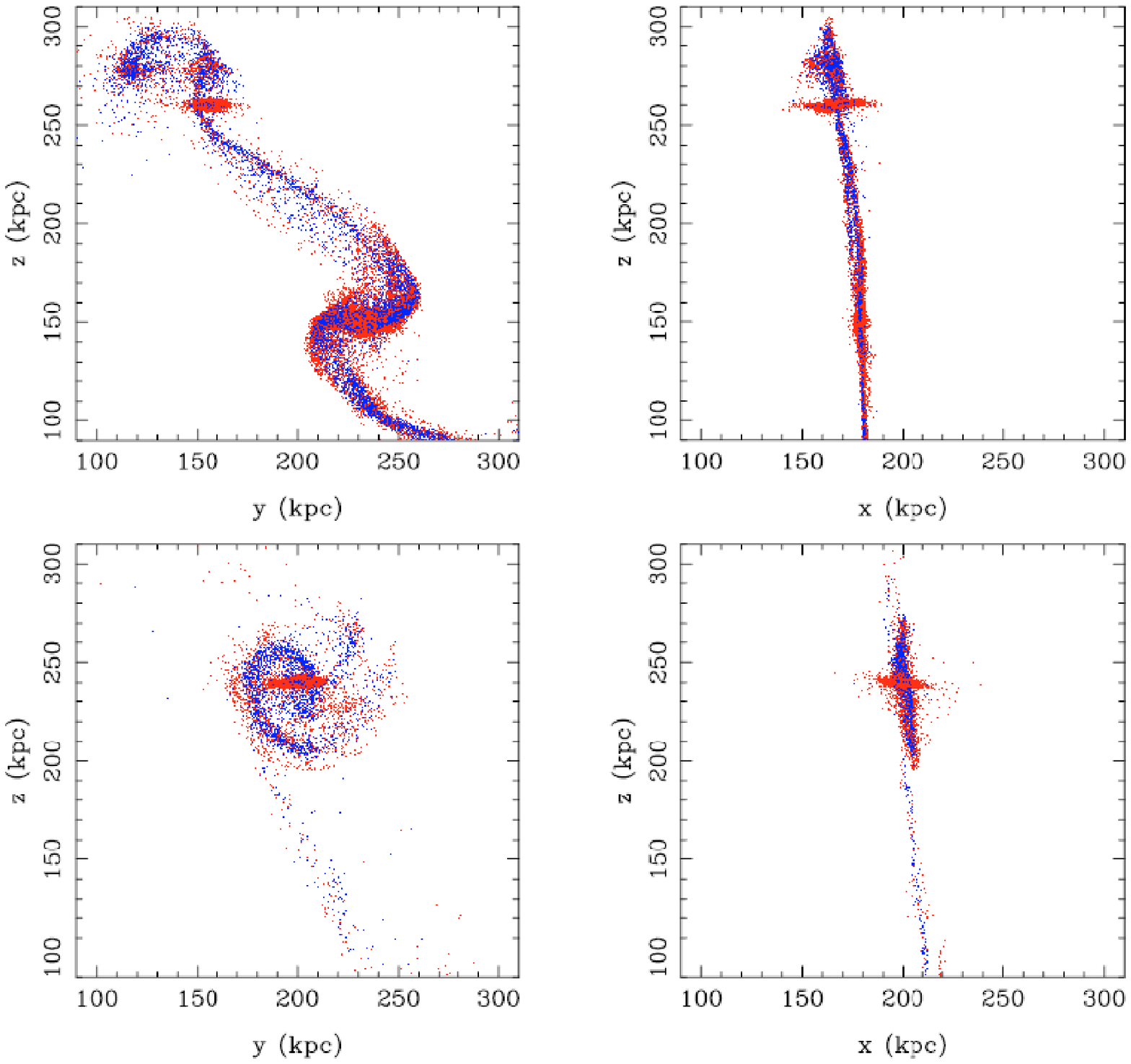}}
\caption{Simulations of the polar ring formation in AM 1934-563,
compared with the observations.
Stellar particles are in red, gas clouds in blue. An incomplete ring
has just formed.
From Reshetnikov \etal (2006).}
\label{fig:AM1934}
\end{figure*}

Some might be even more complex, with two perpendicular PRG,
as in the ESO 474-G26 system (Reshetnikov \etal 2005).
The two perpendicular rings are formed with stars from
each separated companion. In each case, detailed
numerical simulations can constrain the formation scenario 
(cf Figure \ref{fig:AM1934}).

\subsection{Tidal accretion versus merging scenario}

Tidal accretion of gas from a companion flying by in
an hyperbolic orbit has been shown to be more likely
to form the PRG observed (e.g. Bournaud \& Combes 2003).
The process is able to form inclined polar rings,  such as
NGC 660 or NGC 3718, revealing 
intermediate and inclined cases 
suggesting continuity with warps.
When simulating series of different encounters, with
statistics over geometry, initial spin angles, orbital energy, angular
momentum or mass ratios,  the tidal accretion scenario
is 3-5 times more likely than minor mergers to form PRG 
final remnants.
Very massive PR can form through tidal accretion up to 100\% of the host mass. 

In particular, the prototype N4650A appears more a case for tidal accretion,
then minor merger.
There is a test to disentangle the two possibilities:
in case of a minor merger, the stars from the swallowed companion are
dispersed all around, while the gas settles in a polar ring
through dissipation. Then in the final remnant, a halo of
old stars should be detected around the main galaxy. This is not
the case for NGC 4650A, where the polar ring contains
8 10$^9$M$_\odot$ of HI gas and 4 10$^9$M$_\odot$ of young stars,
formed after the event.

Polar ionized gas disks allow the determination of metallicities
and abundances to constrain the formation scenario. In 
IC5181, Pizzella \etal (2013) demonstrate that the polar gas is
likely to come from a donor, the tidal accretion reproducing better
the age, metallicity and [alpha/Fe] data of the object.

The system AM2020-504 involving a rare PRG around an elliptical galaxy,
 favors also the accretion from a donor galaxy and not cold gas accretion.
It is surrounded by a group of nearby galaxies, with about the same velocity.
The ring is young and blue, with relaxed rotation and a warp.
The polar ring follows the metallicity-luminosity relation for normal
spiral galaxies and has an O/H gradient across the ring
(Freitas-Lemes \etal 2012).

AM 1934-563 is part of a triplet galaxy, the primary disk
has a rotational velocity compatible with a spiral galaxy on 
the Tully-Fisher relation. The polar ring has a small mass
and therefore could be an ideal case. The numerical simulations (cf Figure
\ref{fig:AM1934}) favors the tidal accretion scenario from a gas-rich companion
(Reshetnikov \etal 2006).
This scenario was however discussed by Brosch \etal (2007)
with new long-slit spectroscopic data. They 
found the stellar component rotating as a solid body
contrary to the gas component in the primary disk, and the gas in the polar
ring rotating slower that in the primary. They favor the cold accretion
scenario, with dark matter flattened along the primary.

UGC 9796 is a gas-rich PRG, with
5 10$^9$M$_\odot$ of HI in the polar ring. It is surrounded by a group of 
gas-rich galaxies.
The ring is inclined by 25$^\circ$ from polar, and warps away
from the minor axis of the host, as is reproduced in simulations
(Cox \etal 2006).

\subsection{Late gas infall, or dark matter infall with resonance}

Without any interaction or merger, polar rings can form through
gas infalling from cosmic filaments 
(Katz \& Rix 1992, Semelin \& Combes 2005, Brook \etal 2008, 
Snaith \etal 2012).
Contrary to the tidal accretion scenario,
dark matter is also infalling from filaments, in even larger quantity
than gas.
The stability problem, with respect to infalling angles,
is the same as in the merging scenario.

In the cosmological simulations by Maccio \etal (2006),
gas was accreted from cosmic filaments at almost a right angle
from the plane of the host, and all the gas settled in the polar ring,
while stars were only in the host.
Several small companions in the filament were also accreted
as minor mergers. At the end, the global
dark matter halo was quite round in the visible part.

\begin{figure*}[ht]
\centerline{
\includegraphics[angle=-0,width=10cm]{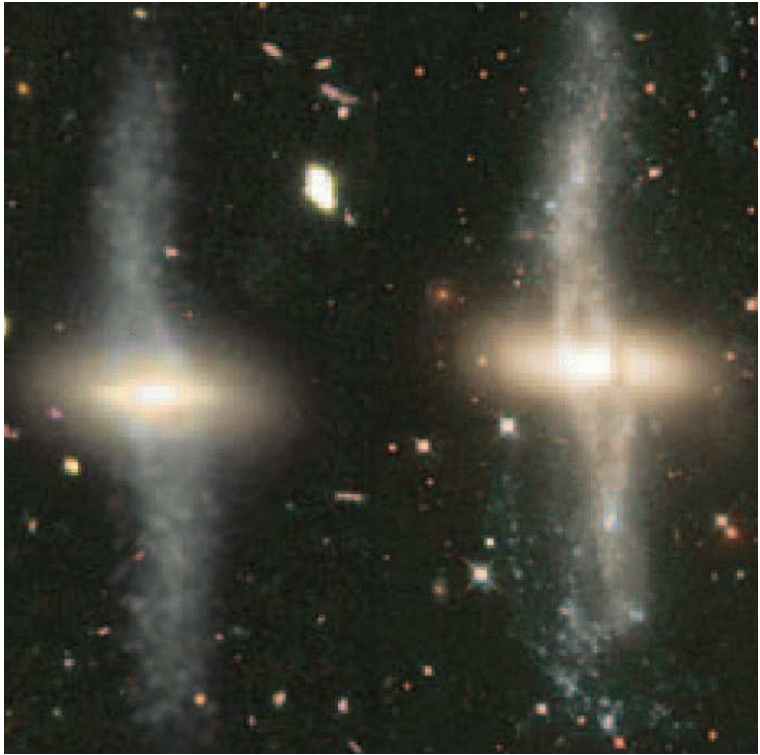}}
\caption{
Composite image of the polar ring simulation and comparison with NGC 4650A
(right) from Gallagher \etal (2002). The simulated
polar disk (left) is imaged by assigning the colors blue, green, and red to three
mock HST bands at 450, 606, and 814 nm, as done in the observations. The final product is superimposed
onto an HST background. 
From Brook \etal (2008).}
\label{fig:N4650A}
\end{figure*}

In the more targetted simulations by Brook \etal (2008),
cf Figure \ref{fig:N4650A}, accretion first formed the host.
After a sudden change of filaments and orientation, accretion
occurs then along the polar system, with both gas and dark matter.
After 1.5 Gyr, the interaction 
between the two disks destroys the PRG.
During the life-time of the PRG, the velocity curve is about the same in both
equatorial and polar planes, contrary to most observations.

The fact that some PRG are polar disks more than polar
rings supports the formation through (misaligned) cosmic accretion
(Spavone \etal 2010), 
while the presence of a true ring supports the tidal accretion.
Snaith \etal (2012) develop in more details the cosmic accretion studied 
 by Brook \etal (2008); they look in particular at the dark matter
when the polar disk is progressively formed out of  matter
infalling from a cosmic filament.
They find that the dark matter is aligned with the polar system.

To be complete, we can also mention another scenario to form PRG,
but which is certainly rarer: the
resonance phenomenon. In a spiral galaxy, embedded
in a triaxial dark matter halo, slightly tilted orbits precess with a 
retrograde slow pattern. They could resonate with the tumbling
retrograde triaxial halo, if the latter does tumble.
When dark matter is accreted, the halo pattern speed tends to zero, and passes
across the resonance, which levitates the orbits in a polar ring.
The polar ring would then contain two equal-mass counter-rotating
stellar disks (Tremaine \& Yu 2000). This mechanism has been invoked
also to explain counter-rotating disks in the equatorial plane
(Evans \& Collett 1994).

\section{Observational tests}
\subsection{Tully-Fisher for PRG}

\begin{figure*}[ht]
\centerline{
\includegraphics[angle=-0,width=6cm]{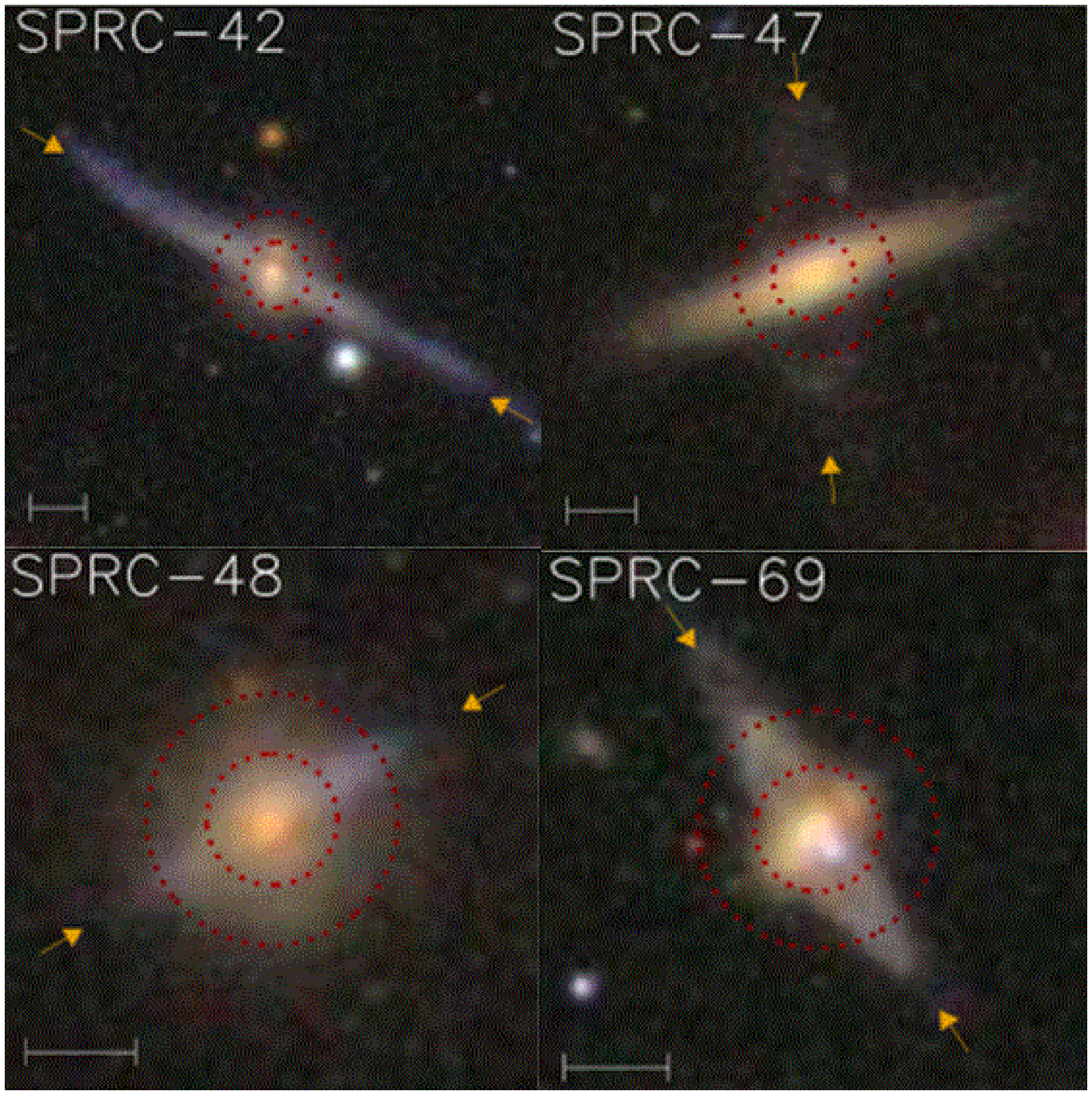}
\includegraphics[angle=-0,width=6cm]{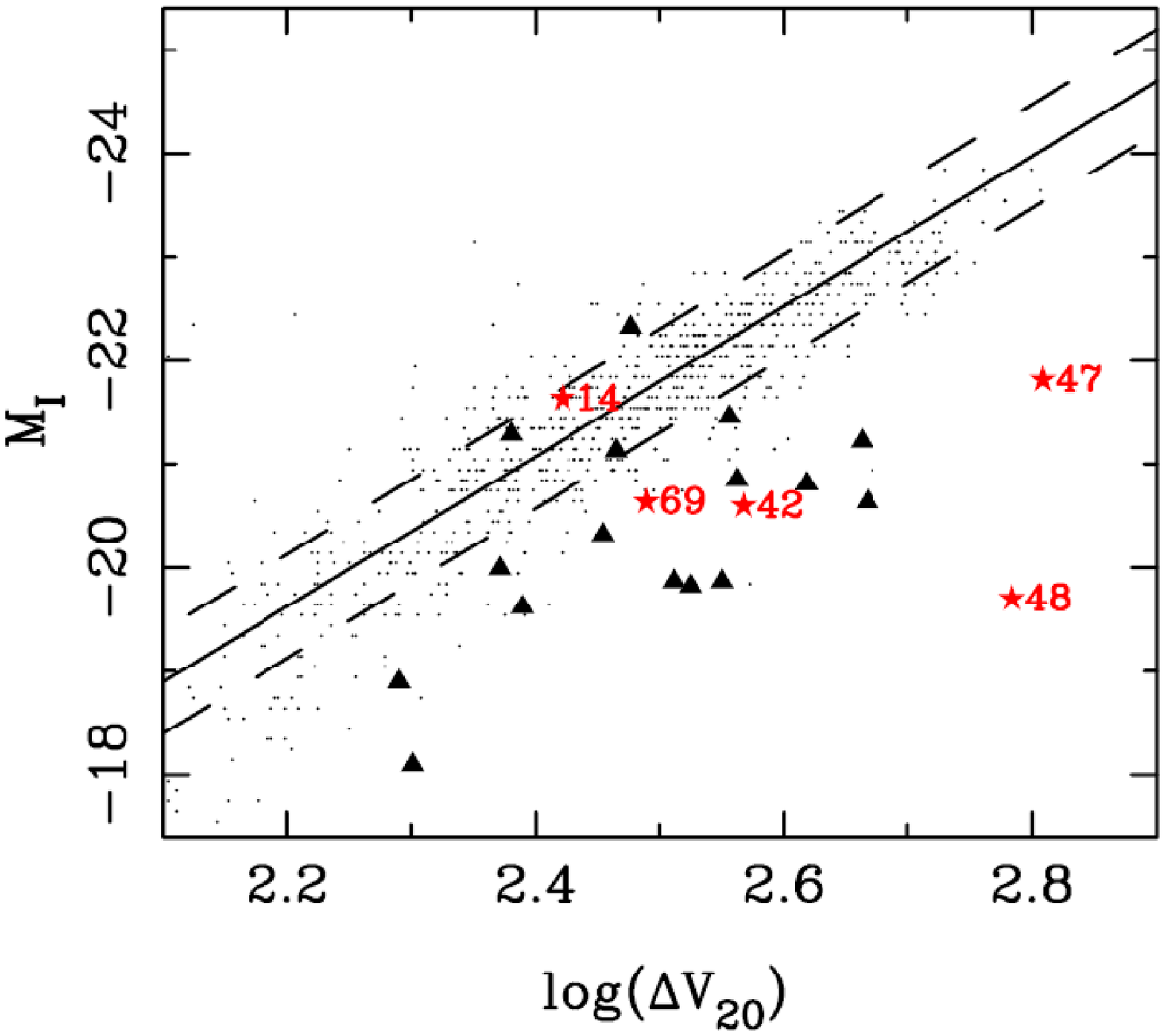}}
\caption{
Molecular gas in polar rings. Left: images of 4 of the detected 
polar systems, with the CO(1-0) and CO(2-1) beam sizes from the IRAM-30m. 
Right: Tully-Fisher relation for our 5 detected PRG (red stars) compared
with other PRG (black triangles, Whitmore \etal 1990, van Driel \etal 2002, Iodice \etal 2003),
and normal spirals (black dots, Giovanelli \etal 1997). The velocities
in the abcissa are the full width at 20\% power obtained from the CO profiles.
From Combes \etal (2013).}
\label{fig:molPRG}
\end{figure*}

The number of individual PRG where detailed kinematics is resolved
is very small. However, it is possible to have more statistics,
with a global velocity measurement, through the HI spectrum of the PR.
More than a dozen have been displayed in a Tully-Fisher (TF) diagram,
like in Figure \ref{fig:molPRG} (Iodice \etal 2003).
The surprise is that most of them reveal a higher rotational
velocity in the polar ring, than expected from their TF location.
While as predicted by simulations, the velocity of the PR should be
lower than in the equatorial plane. Indeed, there is no circular orbits
in a PR system. In addition, the velocity of the PR is measured
when the matter is at apocenter, since in general, for selection effects, 
both components polar and equatorial are seen nearly edge-on. 
When dark matter is spherical or flattened along the host, the
observed velocity for PR should be the smallest. 
And the more dark matter is added in the halo, the 
more excentric are the PR orbits.
Simulations show that the polar rings are indeed significantly elongated perpendicular
to the host potential well, even when massive enough to be 
self-gravitating.

The surprising result could be a consequence of the perturbed host systems, 
in color or luminosity, due to the accretion event. 
However, when the TF of the hosts is compared with the TF of the polar rings, it is possible to rule out this possibility.
The equatorial velocity is indeed lower than the polar velocity.

The main implication of the TF diagram for PRG is that
most of PRG require dark matter aligned along the polar disk,
and not along the host.
Only 2 cases, where the ring is light, can be explained with
the sole visible baryonic mass flattened along the host.
With collisionless dark matter, both merging and tidal accretion scenarios
produce either spherical haloes, or halos flattened along the host.
A secondary accretion event of gas and dark matter is then required.
There could be also some dark baryons, in the form of gas.
Their dissipative character will re-create some halo flattened 
along the PR. The metallicity of the gas could disentangle the scenarios.

Note that an alternative solution to dark matter has been developed
in the form of modified gravity (MOND, e.g. Milgrom 2010). 
Recently, it has been shown that in this model, the
expected rotational velocity in the PR is always larger than in
the host (L\"ughausen \etal 2013).

\subsection{Molecular gas in polar rings}
HI-21cm emission is essential to have PR rotation curves.
But PRG are rare and often at high redshift. Only the
molecular gas could then be detected, and ALMA opens 
large perspectives. 

CO emission has been detected in NGC 660 (Combes \etal 1992), 
in the polar rings of NGC 2685 and NGC 4650A  (Watson \etal 1994),
in the spindle galaxy NGC 2685 with interferometry (Schinnerer 
\& Scoville 2002), or in the inner PR, in the 
 center of NGC2768 (Crocker \etal 2008). In addition,
 ten PRG have a global CO spectrum (Galletta \etal 1997).

We have recently undertaken the CO search in the best candidates
of Moiseev \etal (2011). Five have been detected, and will
be mapped with the IRAM interferometer (Combes \etal 2013). 
Their TF position is indicated in Figure
\ref{fig:molPRG}.

\section{3D-shapes of haloes}

Dark matter haloes could be triaxial. How can we probe
their shape already in the galaxy planes? Tests have
been made of their axisymmetry, with 
HI orbits versus velocity widths (Merrifield 2002).
If the HI orbits were non-circular, more scatter would be expected in the 
Tully-Fisher relation. But this is not observed.
The comparison of the morphology and kinematics of the
ring in IC2006 has also concludeed to axisymmetry (Franx \etal 1994).
Flattening is only expected in the direction perpendicular to the plane.

A first method is the flaring of the HI plane. The ideal
would be to measure simultaneously the shape of the flare (in edge-on
galaxies), and the HI velocity dispersion, perpendicular to the plane
(in face-on galaxies). Since this cannot be done, the HI dispersion
is assumed to be constant, 
$\sigma_z$(HI) $\sim$ 10km/s, as in some face-on galaxies.
Then the shape of the flare in z is a function of the density of dark matter
in the plane. Given the rotation curve, and the total dark matter, 
it is possible to deduce the flattening.
One caveat is to identify all baryons in the plane, 
since some dark baryons can exist there, especially at large radii
(e.g. Kalberla 2004).

\begin{figure*}[ht]
\centerline{
\includegraphics[angle=-90,width=11cm]{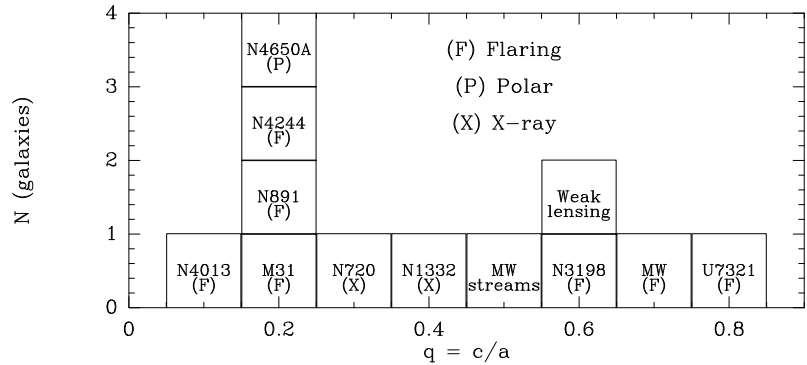}}
\caption{Histogram of halo flattening, according to
the three main methods used in the litterature:
(F) flaring gas-layer, (P) polar-rings and (X) X-ray isophotes. There is also
an estimation from tidal streams in the MW, and a statistical 
galaxy-galaxy lensing estimation.}
\label{fig:flat}
\end{figure*}

The method produced a large diversity of results in the literature,
as shown in Figure \ref{fig:flat}
(O.Brien \etal 2010, Peters \etal 2013). Either
a constant velocity dispersion is assumed, or a model is
designed to derive it. One of the
problems is that the vertical force depends on the halo in the 
extreme outer parts
(cf Olling 1995, Becquaert \& Combes 1997).
Through the HI flaring model,
the dark matter halo in the Milky Way has been derived prolate in shape
(Banerjee \& Jog 2011).

The results of various methods are also compared in Figure \ref{fig:flat}.
The PR method assumes dark matter aligned with the host, and the
HI-flaring method assumies a constant velocity dispersion of 10km/s.
Other methods are the shape of X-ray isophotes,
the use of tidal streams in the Milky Way (Helmi 2004),
or the galaxy-galaxy lensing (see Arnaboldi's review, this meeting).
Whether the halo is truncated or not makes a lot of difference,
it could be truncated at the end of the HI rotation curve, or much farther out 
(Bland-Hawthorn \etal 1997).

Other methods have been proposed in the past, like the
extension of tidal tails to constrain the dark matter shape
and concentration  (Dubinski \etal 1996,
Springel \& White 1999, Dubinski \etal 1999).
Tidal dwarfs can also
constrain the extension of dark matter haloes
(Duc \etal 2004, Bournaud \etal 2003).

\section{Conclusions}

Polar rings are very useful objects to probe the 3rd dimension in dark halo shapes,
but there are caveats.
More statistics are needed, to disentangle all formation scenarios: 
mergers, tidal accretion, cosmic accretion of gas and dark matter.
It is possible that the PRG formation itself modifies considerably 
the dark matter shape. Light polar rings are therefore ideal.
Molecular gas tracers open new perspective, since they 
allow to gather more PRG at high redshifts.
The flaring method is complementary,
however with assumptions on vertical dispersions or on the dark matter radial
 extent. Warps complicate the derivation;
as for polar rings, they can be explained by late accretion of
dark matter and gas.

\acknowledgements Thanks to Enrica Iodice and the organisers for such 
an interesting and nicely located meeting.


\end{document}